\documentclass[aps,prl,10pt,
               preprintnumbers,numbers,sort&compress,
               twocolumn,
               showpacs,showkeys]{revtex4}
\usepackage{graphicx,amsmath,amssymb,bm}
\usepackage{psfrag}

\newcommand{\kf}{k_{\mathrm{F}}}

\newcommand{\ku}{k^\uparrow_{\mathrm{F}}}
\newcommand{\kd}{k^\downarrow_{\mathrm{F}}}

\newcommand{\muu}{\mu^\uparrow}
\newcommand{\mud}{\mu^\downarrow}

\newcommand{\ef}{\varepsilon_{\mathrm{F}}}
\newcommand{\eu}{\varepsilon^\uparrow_{\mathrm{F}}}
\newcommand{\ed}{\varepsilon^\downarrow_{\mathrm{F}}}
\newcommand{\up}{U_{\mathrm{P}}}
\newcommand{\nb}{n_{\mathrm{b}}}
\newcommand{\nf}{n_{\mathrm{F}}}
\newcommand{\ufb}{U_{\mathrm{fb}}}
\newcommand{\ubb}{U_{\mathrm{bb}}}
\newcommand{\afb}{\alpha_{\mathrm{fb}}}
\newcommand{\abb}{\alpha_{\mathrm{bb}}}

\newcommand{\abs}[1]{\lvert{#1}\rvert}
\newcommand{\Abs}[1]{\left\lvert{#1}\right\rvert}
\newcommand{\vect}[1]{\vec{\mathbf{#1}}}

\begin{document}

\title{Induced P-wave Superfluidity in Asymmetric Fermi Gases}
\author{Aurel Bulgac}
\email[E-mail:~]{bulgac@phys.washington.edu}
\author{Michael M$^{\text{c}}$Neil Forbes}
\email[E-mail:~]{mforbes@alum.mit.edu}
\author{Achim Schwenk}
\email[E-mail:~]{schwenk@u.washington.edu}
\affiliation{Department of Physics, University of Washington,
Seattle, WA 98195-1560}


\begin{abstract}
  We show that two new intra-species P-wave superfluid phases appear
  in two-component asymmetric Fermi systems with short-range S-wave
  interactions.  In the BEC limit, phonons of the molecular BEC
  induce P-wave superfluidity in the excess fermions.  In the BCS
  limit, density fluctuations induce P-wave superfluidity in both the
  majority and the minority species.  These phases may be realized in
  experiments with spin-polarized Fermi gases.
\end{abstract}

\maketitle
The phenomenon of fermionic superfluidity spans twenty orders of
magnitude, from cold atomic gases, through liquid $^3$He, electronic
superconductors, nuclei, and neutron stars, to quark color
superconductors.  Experiments with trapped cold atoms provide clean
access to this physics, with tunable interactions and compositions.

In this Letter, we discuss the possibility of \emph{P-wave}
superfluidity in cold asymmetric Fermi gases comprising two species
($\uparrow$ and $\downarrow$) of equal mass $m$, with short-range
\emph{S-wave} interactions.  (The generalization to unequal masses is
straightforward.)  If the system is symmetric (equal number density of
each species), the ground state is well described as a fully gapped
superfluid: There is a crossover from a BCS state at weak attraction
to a BEC of tightly-bound diatomic molecules (dimers) at strong
attraction.

The ground state of asymmetric systems is not even qualitatively well
understood.  If the chemical potential difference $\delta\mu$ between
the two species is larger than the gap $\Delta$ in the spectrum, the
stability of the superfluid state is compromised.  At weak coupling it
is known that there must be a phase transition for $\delta\mu
\leqslant\Delta/\sqrt{2}$ to some asymmetric
state~\cite{Clogston:1962}.  Various possible asymmetric superfluid
phases have been proposed: For example, anisotropic/inhomogeneous
superfluid states with crystalline structure
(LOFF)~\cite{FF:1964-LO:1965} or deformed Fermi
surfaces~\cite{MS:2002}, and homogeneous gapless (breached pair)
superfluids~\cite{Liu:2002gi-Gubankova:2003uj-Forbes:2004cr}.  To
date, all proposed phase diagrams conclude that, for large asymmetry,
one finds a normal Fermi liquid~\cite{PWY:2005,Son:2005qx,SR:2005}.

\begin{figure}[t]
  \psfrag{dmu/Delta0}{$\delta\mu/\Delta_{\delta\mu=0}$}
  \psfrag{-1/ka}{$-1/(a\sqrt[3]{n_{\delta\mu=0}})$}
  \psfrag{1}{1}
  \psfrag{0}{0}
  \psfrag{N1a}{Fully Polarized (one species)}
  \psfrag{N1b}{Fermi Gas}
  \psfrag{P1}{P$_{1}$}
  \psfrag{P2}{P$_{2}$}
  \psfrag{LOFF?}{LOFF?}
  \psfrag{BEC}{BEC}
  \psfrag{BCS}{BCS}
  \begin{center}
    \includegraphics[width=\columnwidth]{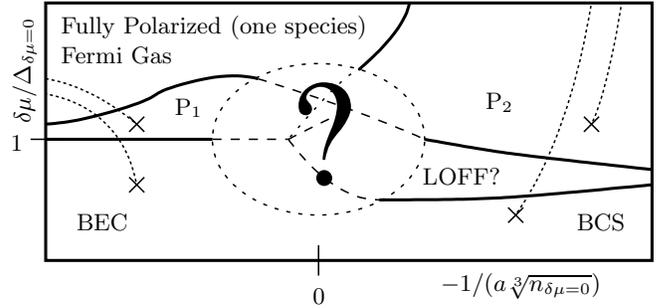}
    \caption{\label{fig:phase} Conjectured zero-temperature grand
      canonical phase diagram after~\cite{Son:2005qx}.  Scaling
      relations allow this two-dimensional projection of the three
      parameters $(\mu,a,\delta\mu)$.  The chemical potential
      asymmetry $\delta\mu = (\muu-\mud)/2$, and the S-wave
      scattering-length $a$ are expressed in units of the S-wave gap
      $\Delta_{\delta\mu=0}$ and density $n_{\delta\mu=0}$ of the
      symmetric state with the same parameters $\mu$ and $a$, but with
      $\delta\mu=0$.  The phase P$_1$---a gapless superfluid with
      single Fermi surface in~\cite{Son:2005qx}---is our P-wave
      superfluid/BEC phase.  The phase P$_2$---a two-component
      Fermi-liquid in~\cite{Son:2005qx}---is our two-component P-wave
      phase.  There may be additional P-wave phases (for instance in
      the breached pair phases) near the S-wave Feshbach resonance.
      Likewise, the exact nature of the phase(s) labeled LOFF?\ has
      yet to be determined.  This phase diagram is illustrative: Our
      results are independent of its quantitative structure.  The
      dotted lines in the BEC and BCS regimes represent sample
      trajectories of constant $a$ and $\delta\mu$.  A trap provides a
      radially varying $\mu$, and would comprise concentric shells of
      the various phases crossed by these trajectories starting from
      the crosses at the center of the trap.  To maximize the physical
      space occupied by these new phases, it might be useful to load
      the trap so the core is a P-wave phase as indicated by the two
      shorter trajectories.  (Note: we have verified the
      qualitative structure and shape of this diagram (outside of the
      unitary regime) with a self-consistent mean-field cross-over
      model.)}
  \end{center}
\end{figure}

Here we show that attractive \emph{intra}-species P-wave interactions
are induced between gapless fermions, leading to the formation of
P-wave superfluids.  We propose a phase diagram (Fig.~\ref{fig:phase})
where phases with excess fermions are P-wave superfluids (except for
the free Fermi gas consisting of a single non-interacting species).
In the BEC regime, the P$_1$ phase consists of a molecular BEC with a
single P-wave superfluid of the excess particles.  In the BCS regime,
the P$_2$ phase has two coexisting P-wave superfluids: one in
the majority component, and one in the minority component.

We calculate the induced interactions that give rise to P-wave
superfluidity in the controlled limit of small scattering length, and
estimate the dependence of the P-wave gap $\Delta_{\text{P}}$ on the
various physical parameters.  These P-wave phases could soon be
observed, as experiments are just starting to explore the properties
of spin-polarized gases~\cite{ZSSK:2005,PLKLH:2005}.
\paragraph{BEC Regime.}
In the BEC limit of strong attraction, the fermions form tightly-bound
molecules that condense into a BEC\@.  For asymmetric gases, the
na\"\i{}ve picture is that the excess fermions interact weakly with
the molecules, and coexist in the ``spaces'' between the molecules.
The resulting phase exhibits gapless fermionic excitations in a
superfluid BEC and has been called a ``gapless superfluid''.  (This
picture is supported by Ref.~\cite{Carlson:2005kg} and mean-field
cross-over models.)  We show, however, that phonon exchange induces an
attractive P-wave interaction between the excess fermions.  This
interaction destabilizes the Fermi surface, forming an
\emph{intra}-species P-wave condensate on top of the molecular BEC\@.

To estimate the magnitude of the induced P-wave interaction $\up$, and
the relevant critical temperature $T_{c}^{\text{P}}$, we consider the exchange
of a single phonon between the excess fermions.  In weak
coupling~\footnote{To leading order in the interaction, the prefactor
  for the angle-averaged gap is $8/e^2$.}, the P-wave superfluid
pairing gap is given by $\Delta_{\text{P}} \sim \ef \,
\exp\bigl[1/(N_{\mathrm{F}} \up)\bigr]$, where $\ef$ is the Fermi
energy of the excess fermions, $N_{\mathrm{F}} = m \kf
/(2\pi^2\hbar^2)$ is the density of states at the Fermi surface, and
$\hbar\kf$ is the Fermi momentum.  The interaction induced by
single phonon exchange is given by~\cite{BBP:1967-VPS:2000}
\begin{equation}
  U_{\mathrm{ind}}(q_0,\vect{q}) =
  \ufb^2 \: \frac{2 \, \nb E_{\vect{q}}}{q_0^2 - E_{\vect{q}} \left( 
      E_{\vect{q}} + 2 \, \nb \ubb \right)} \,,
\end{equation} 
where $\ufb$ is the fermion-boson coupling, $\ubb$ is the boson-boson 
coupling, $E_{\vect{q}}=q^2/(2m_{\mathrm{b}})$, $m_{\mathrm{b}}=2m$ is 
the mass of the molecules, and $\nb$ is the boson density.
The couplings are determined from the scattering length $a > 0$ using 
few-body methods~\cite{STM:1956-PSS:2004-BBF:2003}:
\begin{subequations}
  \begin{align}
    \ufb &= \frac{\pi a\hbar^2}{m} \, \afb 
    \approx \frac{\pi a\hbar^2}{m} \, 3.6 \,, \\
    \ubb &= \frac{\pi a\hbar^2}{m} \, \abb 
    \approx \frac{\pi a\hbar^2}{m} \, 1.2 \,.
  \end{align}
\end{subequations}
We neglect the frequency dependence in weak coupling, and thus the static
induced interaction is
\begin{equation}
  U_{\mathrm{ind}}(0,\vect{q})=
  -\frac{\pi a \hbar^2}{m} \frac{\afb^2}{\abb}
  \biggl[ 1 + \biggl( \frac{q}{2 m_{\mathrm{b}} c} \biggr)^2
  \biggr]^{-1} \,,
\end{equation}
where the speed of sound in the Bose gas is 
\begin{equation}
  c = \sqrt{\nb \, \ubb\,/\,m_\mathrm{b}} = 
  \hbar\,\sqrt{2 \pi a \,\nb\, \abb}\,/\,m_\mathrm{b} \,.
\end{equation}
In weak coupling, the infrared behavior implies
that momenta in the vicinity of the Fermi surface $\abs{\vect{p}_{i}} =
\hbar\kf$ dominate the pairing interaction.  Moreover, the energy transfer
is small, justifying our neglect of the frequency dependence.
Pauli exclusion forbids S-wave induced interactions between the excess
fermions, thus, the induced P-wave interaction is dominant.
Projecting onto the Legendre polynomial $P_1(\cos\theta)$ with $\cos\theta =
1-(\vect{p}_{1}-\vect{p}_{2})^2/(2\kf^2)$ leads to the P-wave
component~\cite{EV:2002}
\begin{subequations}
  \label{eq:Up}
  \begin{align}
    \up &= \int\limits_{-1}^1 \frac{\mathrm{d}\cos\theta}{2} \, \cos\theta \;
    U_\mathrm{ind}(0,\vect{p}_{1}-\vect{p}_{2}) \,, \\
    &= -\frac{\pi a \hbar^2}{m} \frac{\afb^2}{\abb} \, 
    R_{1}\biggl( \frac{\hbar \kf}{m_\mathrm{b} c} \biggr) \,,
  \end{align}
\end{subequations}
where
\begin{equation}
  R_{1}(x) =
 \frac{2}{x^2}\left[
    \left(
      \frac{1}{x^2}+\frac{1}{2}
    \right)\ln(1+x^2)-1
  \right].
\end{equation}
The asymmetry enters through the parameter 
\begin{equation}
  x^2 = \biggl( \frac{\hbar\kf}{m_\mathrm{b} c} \biggr)^2
  = \frac{3 \pi}{\abb \, \kf a} \, \frac{\nf}{\nb} \,,
\end{equation}
where $\nf=\kf^3/(6\pi^2)$ is the density of excess fermions.  The
function $R_{1}(x)$ has a maximum at $R(1.86) =0.104$ and
the following limiting behavior:
\begin{equation}
  R_{1}(x) \rightarrow \begin{cases}
    x^2/6& \text{where } x \ll 1 \,,\\
    \ln(x^2)/x^2 & \text{where } x \gg 1 \,.
  \end{cases}
\end{equation}
For small asymmetries, $\nf/\nb \ll \kf a \ll 1$, we find the
intriguing result that the P-wave gap is independent of the scattering 
length,
\begin{equation}
  \frac{\Delta_{\text{P}}}{\ef} 
  \sim \exp\biggl(
    - \frac{4 \abb^2}{\afb^2} \frac{\nb}{\nf}
  \biggr) 
  \sim \exp(-0.44 \, \nb/\nf) \,,
\end{equation}
although it is suppressed for small asymmetry.  For large asymmetries,
$\nf/\nb \gg \kf a$, the P-wave gap is
\begin{equation}
  \frac{\Delta_{\text{P}}}{\ef} 
  \sim 
  \exp\biggl(
    - \frac{6 \pi^2}{\afb^2 \, (\kf a)^2 \ln(x^2)} \frac{\nf}{\nb}
  \biggr) \,.
\end{equation}
If we fix the interaction and density of excess fermions (constant $a$
and $\kf$), we find a maximal P-wave gap as a function of the boson
density:
\begin{equation}
  \label{eq:P1max}
  \frac{\Delta_{\text{P}}^\text{max}}{\ef}\,
  \sim \, \exp\biggl(
  - \frac{19.2\, \pi \, \abb}{\kf a\, \afb^2}
  \biggr) 
  \sim \, 
  \exp\biggl(
  -\frac{5.6}{\kf a}
  \biggr),
\end{equation}
for a small intermediate asymmetry:
\begin{equation}
  \label{eq:n1max}
  \nf/\nb \approx 0.44 \, \kf a \ll 1 \,.
\end{equation}
Note that this maximal P-wave gap depends parametrically on $1/(\kf a)$
in the exponent as for S-wave pairing in the BCS regime.  Therefore,
we expect similar high critical temperatures towards resonance.

It is instructive to compare the critical temperature for P-wave
superfluidity $T_c^\text{P} \sim \Delta_{\text{P}}$ to the critical
temperature for Bose condensation of the molecules $T_c^{\text{BEC}}
\sim \nb^{2/3}/(\zeta(3/2)^{2/3}m_{\mathrm{b}})$.  For the maximum
P-wave gap, we find, with Eq.~(\ref{eq:P1max}) and~(\ref{eq:n1max}),
the ratio
\begin{equation}
  \frac{T_c^{\text{P, max}}}{T_c^\text{BEC}} \sim 
  (\kf a)^{2/3} \, \exp\left(-\frac{5.6}{\kf a} \right) \,,
\end{equation}
which increases toward resonance.  (Neglecting higher order effects,
the prefactor may be calculated and is of order unity.)  We therefore
conclude that the P-wave superfluid/BEC phase P$_1$ may be observed in
asymmetric Fermi gases as one approaches the S-wave Feshbach resonance
from the BEC regime.

Our results are justified in weak coupling. The omitted pre-exponential
factors, however, depend on higher-order induced interactions. It is
expected that the resulting factors are of order unity.
We also note that the pairing interaction
induced by single phonon exchange is related to the effective mass
$m^*/m=1+F_1/3$ through the Landau parameter $F_1=- N_{\mathrm{F}}
\up$.  Effective mass corrections are thus higher order, but they
would increase the density of states at the Fermi surface, and thus 
increase the magnitude of the P-wave gap and $T^\text{P}_c$.

\paragraph{BCS Regime.}
Next, we show that all proposed asymmetric Fermi liquid phases are 
unstable towards a two component P-wave superfluid due to the exchange
of density fluctuations. This
occurs in the BCS regime and is denoted by P$_2$ in our phase diagram
Fig.~\ref{fig:phase}.  We start from a two component asymmetric Fermi
gas with Fermi momenta $\ku > \kd$, and calculate the induced
interactions in weak coupling.  To lowest order in the S-wave
interaction $4 \pi a \hbar^2/m$, the induced interaction for 
back-to-back scattering is given 
by~\footnote{For pairing in the $l$-wave, the exchange contribution is
$-(1+{\bm \sigma}_1 {\bm \sigma}_2)/2 \, 
U_\mathrm{ind}^{\uparrow\uparrow}(0,\vect{p}_1+\vect{p}_2) \to - 
(-1)^{l+S+1} \, U_l$, which is implicitly included when solving a gap 
equation in the relevant partial wave.}:
\begin{multline}
  U_\mathrm{ind}^{\uparrow\uparrow}(0,\vect{p}_1-\vect{p}_2) = 
  \parbox{2mm}{\includegraphics[scale=0.725]{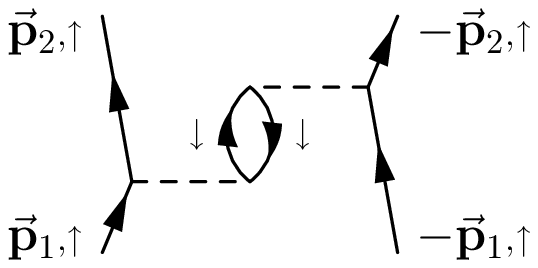}}\\[2mm]
  = - N_\mathrm{F}^\downarrow \, \biggl( \frac{4 \pi a \hbar^2}{m} \biggr)^2
  L\bigl(\abs{\vect{p}_{1}-\vect{p}_{2}} / (2\hbar\kd)\bigr) \,.
\end{multline}
The induced interaction for the minority fermions is obtained by
interchanging the spin labels.  As before, in weak coupling we neglect the 
frequency dependence and consider momenta on the
Fermi surface.  Thus, $L(y)$ denotes the static Lindhard function
\begin{equation}
  L(y)=\frac{1}{2}+\frac{1-y^2}{4y} \ln\Abs{\frac{1+y}{1-y}} \,.
\end{equation}
The importance of induced interactions for superfluidity has been
pointed out for symmetric Fermi systems: in weak coupling for S-wave
pairing~\cite{GMB:1961-HPSV:2000}, for P-wave pairing with repulsive
interaction~\cite{EMBK:2000}, and close to the Feshbach
resonance~\cite{GJB:2005} (based on~\cite{BB:1973}).  In addition, it
has been shown that induced interactions significantly suppress the
superfluid gaps in neutron stars~\cite{LS:2000-SFB:2002,SF:2003}.
For P-wave pairing in neutron stars, it is known that central induced
interactions are attractive~\cite{PR:1991}, but repulsive
spin-orbit fluctuations 
dominate this effect~\cite{SF:2003}.

The resulting P-wave superfluid gap for the majority component is
given by $\Delta_{\text{P}}^\uparrow \sim \eu \,
\exp\bigl[1/(N_\mathrm{F}^\uparrow \up^{\uparrow\uparrow})\bigr]$,
where $U_p^{\uparrow\uparrow}$ denotes the P-wave projection of the
induced interaction as in Eq.~(\ref{eq:Up})~\cite{Kagan}. This leads to
\begin{equation}
  \frac{\Delta_{\text{P}}^\uparrow}{\eu} \, 
  \sim \, \exp\biggl(-\frac{\pi^2}{4 \, 
    \ku \, \kd \, a^2 L_{1}(\ku/\kd)}\biggr) \,,
\end{equation}
with the P-wave superfluid gap for the minority component given by
interchanging the spin labels. The asymmetry enters through the
function
\begin{equation*}
  L_{1}(z) = \frac{5z^2-2}{15z^4}\ln\Abs{1-z^2}
  -\frac{z^2+5}{30z}\ln\Abs{\frac{1-z}{1+z}}
  -\frac{z^2+2}{15z^2} \,,
\end{equation*}
which has the limiting behavior
\begin{equation*}
  L_{1}(z) \rightarrow \begin{cases}
    z^2/18& \text{where } z \ll 1 \,,\\
    \kappa + (7-4\ln2)(z-1)/15 & \text{where } z \approx 1 \,,\\
    2\ln(z)/(3 z^2) & \text{where } z \gg 1 \,.
  \end{cases}
\end{equation*}
For the symmetric case, we recover the result of~\cite{EMBK:2000},
$\Delta_{\text{P}} \sim \ef\,\exp[-\pi^2/(4\,\kf^2 a^2 \kappa)]$, with
$\kappa=(2\ln{2}-1)/5$, however, the work of~\cite{EMBK:2000}
considered repulsive S-wave interactions.  In our case, inter-species
S-wave pairing will dominate for the symmetric system.  The phase
P$_2$ will start for some small but finite asymmetry, and the
deviations in the exponent will be linear in $(z-1)$.

For large asymmetries $\ku \gg \kd$, the P-wave gap of the majority
component is
\begin{equation}
  \frac{\Delta_{\text{P}}^\uparrow}{\eu} \, 
  \sim \, \exp\biggl(-\frac{3\pi^2}{2 \,
    (2 \, \kd a)^2 \ln(\ku/\kd)} \frac{\ku}{\kd} \, \biggr) \,,
\end{equation}
while that of the minority component is
\begin{equation}
  \frac{\Delta_{\text{P}}^\downarrow}{\ed} \, 
  \sim \, \exp\biggl(
  -\frac{18\pi^2}{
    (2 \, \kd a)^2}
  \frac{\ku}{\kd} \, \biggr) \,.
\end{equation}
The majority component has a larger gap, but both are suppressed for
large asymmetry.

For fixed $\kd$, the minority gap $\Delta_{\text{P}}^{\downarrow}$
decreases monotonically for increasing asymmetry, while for fixed
$\ku$, the majority gap $\Delta_{\text{P}}^{\uparrow}$ has a maximum at $\kd
\approx 0.77 \, \ku$, due to the maximum of $L_{1}(z)/z=0.11$ for $z=1.3$:
\begin{equation}
  \label{eq:P2max}
  \frac{\Delta_{\text{P}}^{\uparrow, \mathrm{max}}}{\eu}\,
  \sim \, 
  \exp\biggl(-\frac{\pi^2}{0.11 \, (2 \, \ku a)^2} \biggr) \,.
\end{equation}
Finally, we note that the P$_2$ phase does not destabilize LOFF, or
similar phases, whose condensation energy is parametrically the same
as that of the S-wave BCS phase where $\Delta_\text{S} \sim
\exp(\pi/2 \, \kf a)$.  Thus, the P-wave energy gain is parametrically
smaller in weak coupling.

\paragraph{Discussion.}
Several asymmetric phases proposed in the literature contain Fermi
surfaces, including the normal Fermi liquid phases as well as the
gapless breached pair phases.  Kohn and Luttinger~\cite{KL:1965}
pointed out that, at zero temperature, \emph{all} Fermi surfaces are
unstable in the presence of interactions.  We have shown that, in weak
coupling, induced interactions lead to the formation of P-wave
superfluids with maximal gaps for intermediate asymmetries.  Thus, the
suggested normal Fermi-liquid phases and breached pair phases are
replaced by P-wave superfluids at zero temperature.  In the asymptotic
regions (both in asymmetry and coupling), the gaps are exponentially
suppressed.  The thermodynamic properties of these phases will thus
still be dominated by their underlying Fermi-liquid/breached pair
nature.  However, the dependence of the P-wave gaps on the S-wave scattering
length suggests that the induced P-wave superfluids may be
observable as one approaches the Feshbach resonance, in particular from
the BEC regime.

The nature of both phases P$_{1,2}$ is that of P-wave pairing between
fermions of the \emph{same} species with order parameter
$\langle\mathbf{\hat{a}}_{\vect{p}} \, \mathbf{\hat{a}}_{-\vect{p}}\rangle \sim
\Delta_{\text{P}}(\vect{p}) =
\sum_{m}\Delta_{m}Y_{1,m}(\Omega_{\vect{p}})$.  There are two
qualitatively different states: one with two nodal points
($\Delta_{\text{P}} \sim Y_{1,\pm1 }$) that breaks time-reversal
invariance, and one with a nodal plane ($\Delta_{\text{P}} \sim
Y_{1,0}$).  Energetic calculations suggest that the state with broken time-reversal is favored~\cite{AM:1961}.  Both states spontaneously
break rotational invariance, so the low-energy spectrum will include a
new Nambu-Goldstone boson~\cite{Bedaque:2003wj} in addition to the
gapless nodes.  In the P$_2$ phase, there may be interesting
entrainment physics, since the two P-wave superfluids are coupled by
the S-wave interaction.

As shown in Fig.~\ref{fig:phase}, the phases P$_{1,2}$ will occupy
shells in an optical trap at sufficiently low temperatures.  As long
as there is a finite molecular component $\nb$, a thin shell of the
P$_1$ phase will always be realized.  (The physical volume may be
maximized by tuning the densities.)  Observations of P$_1$ may also be
facilitated by its occurrence in the BEC regime.  The P$_2$ phase will
occupy a large physical volume in the outer regions of the cloud
containing an asymmetry in the BCS regime.  One mechanism for
detecting these phases would be to look for rotational asymmetries in
the momentum distribution or pairwise
correlations~\cite{ADL:2004-GRSJ:2005} in the time-of-flight expansion
images of a single species.  An interesting question is how the trap
asymmetries would influence the spontaneously chosen orientation of
the P-wave state.

\acknowledgments 
We would especially like to thank George Bertsch for discussions
concerning the P$_2$ phase.  We also thank Anton Andreev and Dam Son
for useful discussions.  This work was supported by the US Department 
of Energy under Grant DE-FG02-97ER41014.

\end{document}